\documentclass[9pt,twocolumn,twoside]{revtex4}

\usepackage{siunitx}
\usepackage{hyperref}
\usepackage{amsmath}
\usepackage{amssymb,amsfonts,textcomp}
\usepackage{graphicx}
\newcounter{ExtendedDataFigure}

\usepackage{placeins}

\usepackage{flushend}

\begin{document}

\title{Continuous Wave Multipass Microscopy}
\author{Brannon B. Klopfer}
\affiliation{Applied Physics Department, Stanford University, Stanford, California 94305, USA}
\email{bklopfer@stanford.edu}
\author{Mark A. Kasevich}
\affiliation{Physics Department, Stanford University, Stanford, California 94305, USA}

\begin{abstract}
We present a continuous-wave, post selection-free implementation of a widefield optical multipass microscope.
It can be operated with a spatially and temporally incoherent light source, and requires no active outcoupling or exotic detection schemes.
This implementation is capable of deterministically interrogating a sample sequentially up to $m=4$ times.
Through multiple interrogations, a linear enhancement in phase shift and absorption imparted by the sample can be achieved, fundamentally increasing the signal-to-noise of the obtainable images.
\end{abstract}

\maketitle

\section{Introduction}
In label-free optical microscopies, the phase shift and absorption cross-section of the image target set fundamental and practical limits on microscope performance given the nature of shot noise and the finite performance of optical detectors \cite{adimec_qpi,welldepth_adimec,fluor_snr,snr_is_important}.
In principle, a multipass imaging protocol \cite{mpm,oam,theoryslog,thomasqem,marianqem,qemdesign} --- one in which the optical probe field transits the image target multiple times --- can enhance performance by increasing the phase shift and absorption of the probe field.
For a weakly absorbing sample, this increase means that substantially fewer photons are required to achieve equivalent image quality (as quantified by the image signal-to-noise ratio) for images whose noise is limited by the statistical noise of the probe field (photon shot-noise) or other technical noise sources \cite{theoryslog, thomasqem}.
For example, in an ideal four-pass implementation limited by photon shot noise, the absorption and therefore image contrast increases by a factor of four, resulting in a 16-fold decrease in the number of photons needed to achieve a detection signal-to-noise ratio equivalent to a conventional single pass instrument.
This enables, for example, low-illumination imaging of damage-sensitive samples, or reduced signal acquisition times for dynamic samples.
Likewise, higher image SNR can be achieved with equivalent numbers of illuminating photons, as the image SNR of a single acquisition frame improves linearly with the number of interrogations.
The gains obtainable with this classical method appear to compete favorably with those recently demonstrated with quantum light sources \cite{casacioQuantumenhancedNonlinearMicroscopy2021}.

\begin{figure*}[!htbp]
	\centering
	{\includegraphics[width=\textwidth]{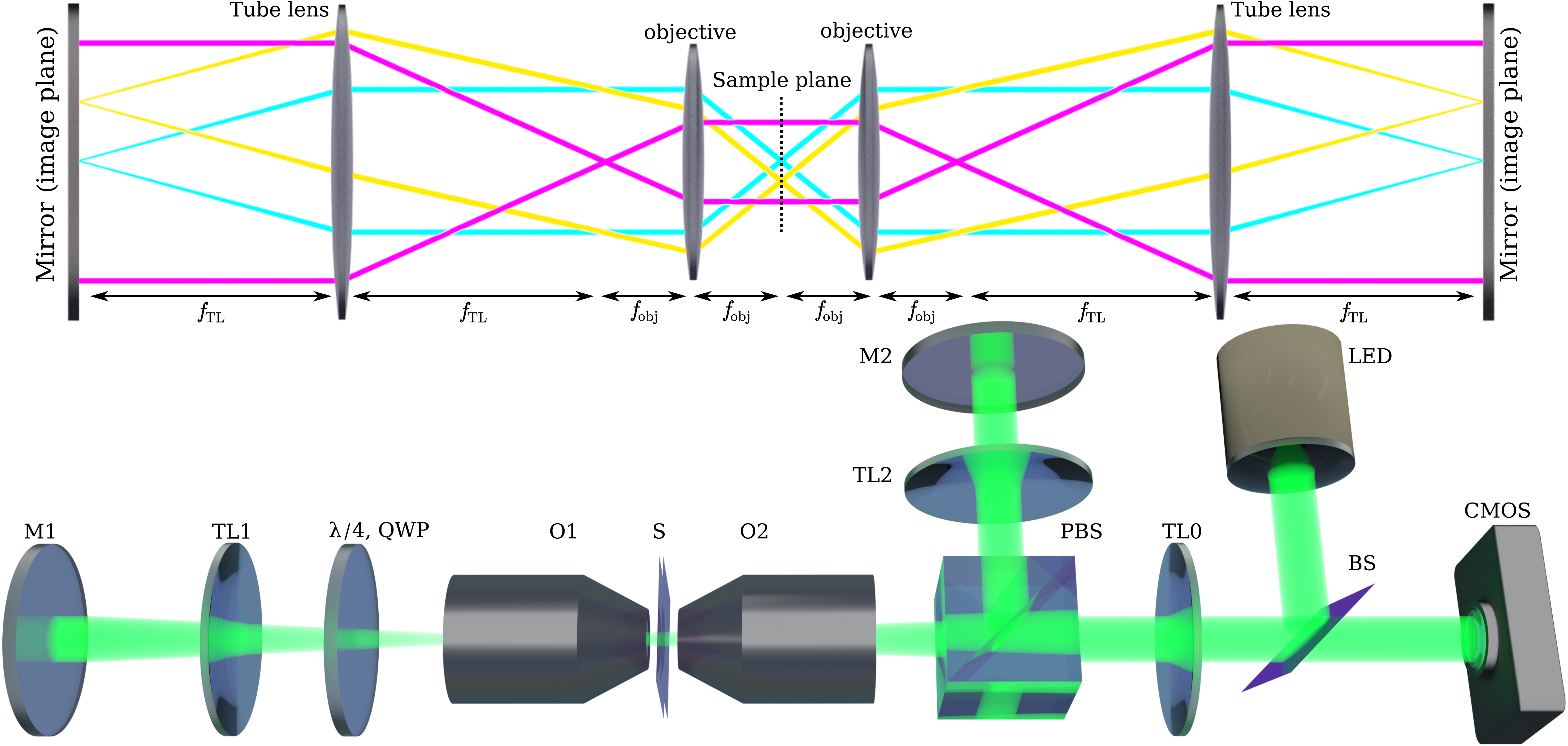}}
	\caption[setup]{
	{\bf Top:}
	ray trace of a multipass microscope with tube lens and objective focal lengths $f_\text{TL}$, $f_\text{obj}$, respectively.
	Rays are shown for plane wave illumination (magenta), on-axis point scatterer (cyan), and off-axis point scatterer (yellow).
	{\bf Bottom:}
	optical setup, consisting of two microscopes placed between end mirrors M1[2].
	The microscopes consist of an NA$=0.8$ objective (O1[2]) and a tube lens (TL1[2]).
	The light (LED) is in- and outcoupled via a beamsplitter (BS) and a polarizing beamsplitter (PBS).
	A quarter-waveplate ($\lambda/4$, QWP) is used to switch between $m=2$ and $4$ interrogations.
	$m = 1$ is obtained by severely defocusing the left-hand objective O1.
	For clarity, additional extracavity optics are omitted.
	}
	\label{fig:setup}
\end{figure*}

In this work, we describe a wide-field multipass implementation compatible with continuous-wave (CW) illumination and conventional cameras.
The implementation can accommodate spatially and temporally incoherent sources, and can achieve up to four passes through the sample.  While scanned-cavity methods such as those described in Refs. \cite{motschCavityenhancedRayleighScattering2010,changCavityQEDAtomic2012} can achieve significantly greater signal enhancements, they cannot operate in  wide-field or incoherent illumination modalities.  Wide-field methods enable substantially faster image acquisition times and incoherent illumination suppresses spurious contributions from scatterers out of the focal plane.

\newcommand{\Ei}{\ensuremath{E_\text{i}}}
\newcommand{\Es}{\ensuremath{E_\text{s}}}
\newcommand{\Is}{\ensuremath{I_\text{s}}}
\newcommand{\Er}{\ensuremath{E_\text{r}}}
\newcommand{\Ir}{\ensuremath{I_\text{r}}}
\newcommand{\Id}{\ensuremath{I_\text{d}}}
\newcommand{\tavg}[1]{\overline{#1}}

Previous demonstrations of multipass protocols relied on pulsed illumination and time-gated cameras to form the multipass image \cite{mpm,oam}.
This approach introduced fundamental inefficiencies in the imaging process by detecting only a fraction of the light incident on the sample, thus failing to fully realize possible imaging gains \cite{theoryslog}.

\section{Background}
It is useful to describe the imaging signal-to-noise and contrast in terms of the reference (unscattered) and scattered components.
For brightfield imaging, we have
\begin{eqnarray}\label{eq:imaging}
\Id \sim \left|\tavg{\Er} + \tavg{\Es}\right|^2 \sim \Ir + \Is + 2 \Er \Es \cos{\phi},
\end{eqnarray}
where \Id{} is the detected intensity, \Er{} (\Ir) and \Es{} (\Is) are the reference and scattered field (intensity), and $\phi$ is the phase shift between the reference and scattered fields \cite{iscat_chapter}.

For coherent illumination in the shot-noise limit, the noise on the reference beam scales as the square root of \Ir{} and thus linearly with \Er.
Taking the signal to be the cross term in \autoref{eq:imaging}, we can estimate the signal-to-noise scaling as
\begin{eqnarray}\label{eq:snr}
\text{SNR} \sim \frac{2 \Er \Es \cos{\phi}}{\Er} = 2\Es\cos{\phi}.
\end{eqnarray}
In the limit of weak samples, the scattered field is much smaller than the reference field, thus the scattered intensity $\Is \sim \Es^2$ can be neglected.
This demonstrates that, in this limit, the fundamental image signal-to-noise is independent of the strength of the reference, and only depends on the strength of the scattered field.
The image contrast $c$, on the other hand, is given by the ratio of the signal to background, and depends on both the strength of the scattered and reference fields,
\begin{eqnarray}\label{eq:contrast}
c \sim \frac{2 \Er \Es \cos{\phi}}{\Ir} \sim \frac{2 \Es \cos{\phi}}{\Er}.
\end{eqnarray}
Methods that improve contrast by reducing the strength of \Er{} (by attenuating this field in the detection path) do not fundamentally change the image signal-to-noise.
Such is the case for iSCAT \cite{iscat-overview} and related techniques, \emph{e.g.}, Fourier filtering \cite{pupil_engineering}, twilight-field microscopy \cite{twilight}, and certain implementations of Zernike phase contrast \cite{zernike_og, zernike_nobel}.
On the other hand, the multipass method fundamentally improves image signal-to-noise \cite{theoryslog,qoptimal1,qoptimal2}.

\begin{figure*}[!htbp]
	\centering
	{\includegraphics[width=\textwidth]{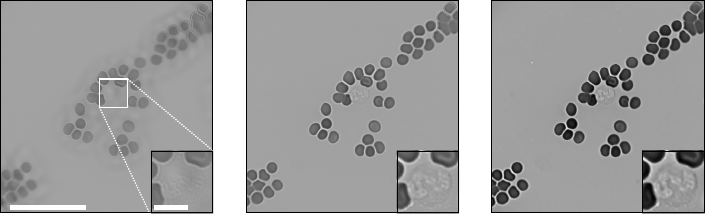}}
	\caption[HRBC]{
Brightfield micrographs of human blood cells at (left to right) $m=1$, 2, and 4, showcasing a wide field of view.
Scale bar is \SI{50}{\micro\metre}.
The halos visible at $m=1$ are due to operating the apparatus in a severely defocused $m=2$ configuration (see text).
Inset shows a white blood cell.
Scale bar on inset is \SI{10}{\micro\metre}.
	}
	\label{fig:widefield}
\end{figure*}

\section{Methods}
The basic CW multipass configuration is illustrated in \autoref{fig:setup}.  In this implementation, the imaging beam transits a self-imaging optics path either 2 or 4 times, depending on the optics configuration.

For a two-pass implementation, illumination light is incoupled through a non-polarizing beamsplitter (BS).  The illumination field is focused in the sample region with objective O2.
After transiting the sample, objective O1 collects the unscattered illumination field and and the scattered image field.
These fields are then imaged by a tube lens TL1 onto a reflective planar mirror M1.
By symmetry, this reflected image is reimaged onto the sample, when appropriately aligned.
Ray-traces of the relevant optical paths are also shown in \autoref{fig:setup}. In order to tune the focus, both the O1 objective and sample stage are actuated along the optical axis with encoded piezo stages.

A four-pass implementation utilizes a polarizing beam splitting cube (PBS) and a quarter-wave plate ($\lambda/4$, QWP).
In this case, the linear polarization of the beam is rotated by $\pi/2$ as it passes twice through QWP, resulting in redirection of the retroreflected field (after having passed twice through the sample) onto a planar mirror M2.  This mirror reflects the fields back into the microscope optical path as in the two-pass implementation.  After two more passes through the QWP (four total passes through the sample), the beam exits the microscope into the image forming optics.  We toggle between the two-pass and four-pass configurations by adjusting QWP: rotating the polarization of the retroreflection leads to 4 passes, while not rotating it leads to 2 passes.

To operate the apparatus in an $m=1$ configuration (\emph{i.e.} without multipassing the sample), we operate in the $m=2$ configuration and severely defocus the objective O1 (see \autoref{fig:setup}).
In this configuration the sample is re-illuminated with a defocused image of the sample, with only the other objective O2 forming an image on the camera.
This was done such that images at $m=1,2$ and $4$ used identical optics and light sources.
As such, a defocused halo can be observed at $m=1$ for certain samples.

\begin{figure}[ht]
	\centering
	{\includegraphics[width=\columnwidth]{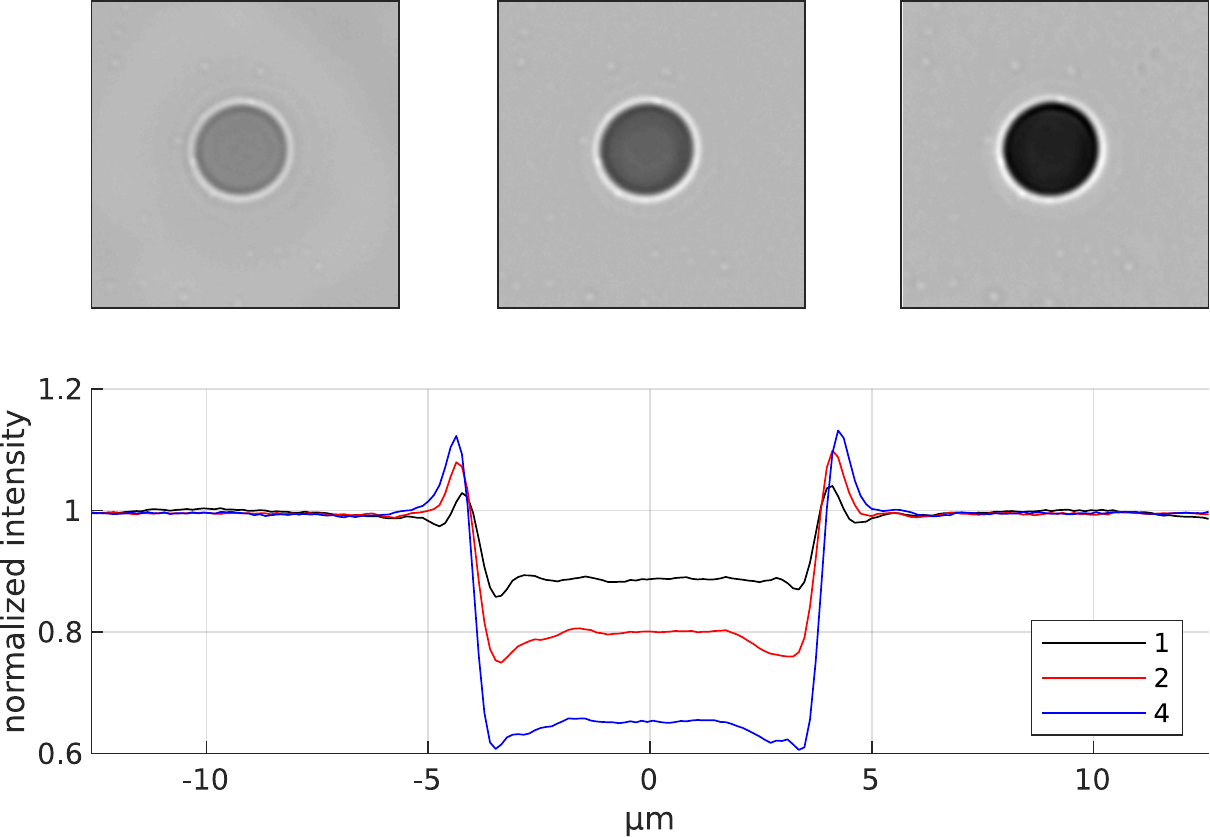}}
	\caption[HRBC]{
	{\bf Top:}
		Micrographs of a human red blood cell (HRBC) at (left to right) $m=1$, 2, and 4.
		The images are normalized to the illumination light.
		Field of view is $\SI{25}{\micro\metre} \times \SI{25}{\micro\metre}$.
	{\bf Bottom:}
		Lateral cross sections.
		Note the dramatic increase in modulation depth as a function of interrogation.
	}
	\label{fig:hrbc}
\end{figure}

\section{Results \& Discussion}
Multipass images of human blood cells at $m=1$, 2, and 4 are shown in \autoref{fig:widefield}.
The contrast enhancement is readily visible (the color scales are held constant for all images).
A human white blood cell (HWBC) is shown in the inset, showing contrast enhancement with a more complex sample.

An individual human red blood cell (HRBC) is shown in \autoref{fig:hrbc}, along with lateral cross sections.
The cross sections display roughly linear buildup in modulation depth as a function of the number of interrogations, as expected.

We estimate the noise of our apparatus by taking the standard deviation per pixel from the central $100\times\SI{100}{px}$ of an image with no sample.
We find the background intensity fluctuations, as normalized to the detected intensity, to be $10^{-3}$ per pixel.
The signal-to-noise may alternately be taken from a time series, which yields the same result after adjusting for common mode intensity fluctuations.
This low spatiotemporal noise value results from both the use of a deep well camera (Adimec Q-2HFW, \SI{2000}{kel} well depth) and incoherent LED illumination (Thorlabs M530F2).
This suggests, for example, that detection of a \SI{100}{\um^2} absorber is possible with a sensitivity to absorption better than $\delta \alpha/\alpha = 4\times10^{-6}$ in a single frame for $m=4$ with the camera used in this work.
For a restricted ROI of $64\times\SI{64}{px}$, these images can be acquired at a frame rate of over \SI{7000}{frames/s}, or \SI{549}{frames/s} for a full field of view ($1440\times\SI{1440}{px}$).
High sensitivity, high throughput dynamic absorption measurements have applications in neuroscience \cite{alfonsoLabelfreeOpticalDetection2020} and flow cytometry \cite{fast_flow00,fast_flow01,fast_flow02}.

\begin{figure}
	\centering
	{\includegraphics[width=\columnwidth]{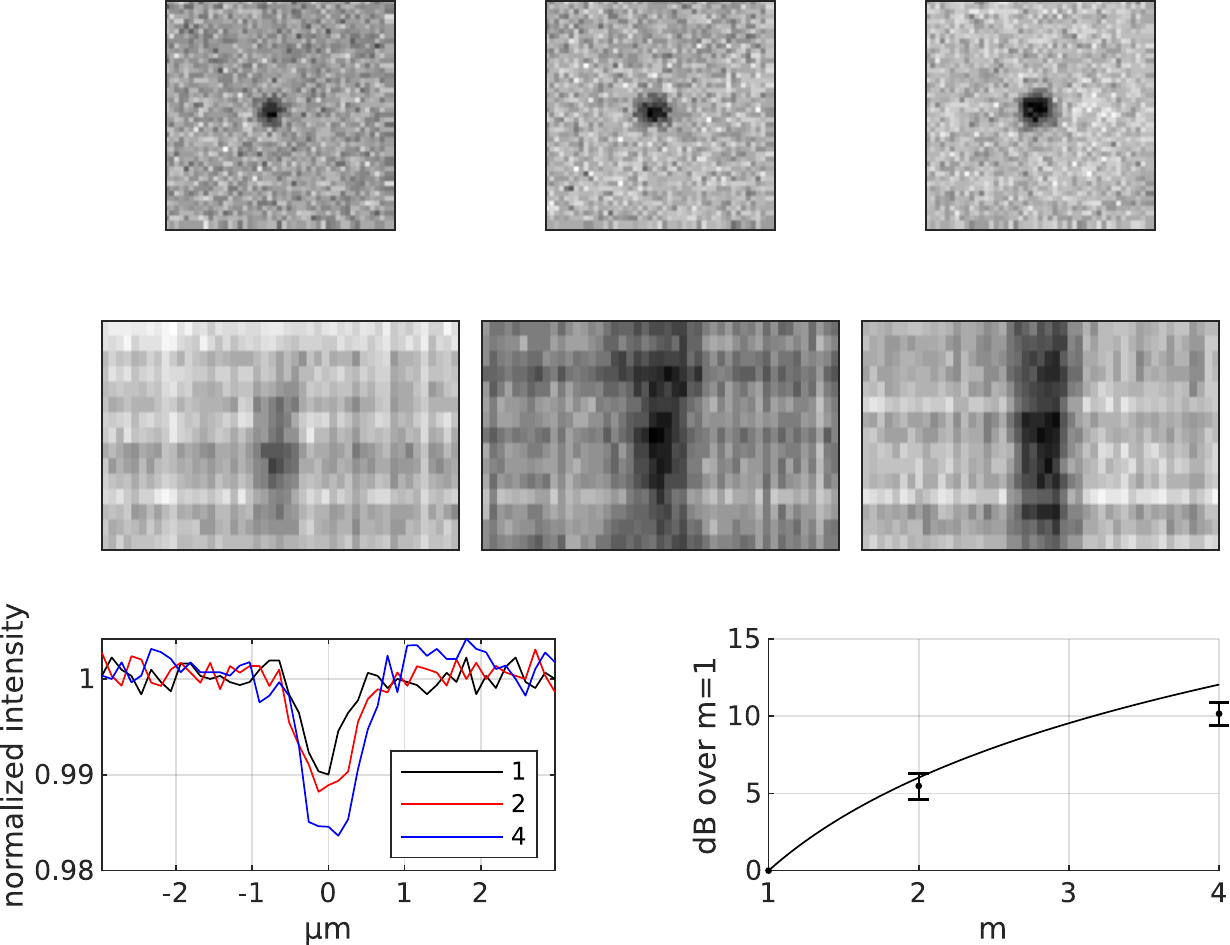}}
	\caption[Gold nanoparticle]{
	Micrographs of a \SI{40}{\nm} gold nanoparticle at (left to right) $m=1$, 2, and 4.
	{\bf Top:}
		real-space brightfield images.
		Field of view is $\SI{6}{\micro\metre} \times \SI{6}{\micro\metre}$.
	{\bf Middle:}
		Cross sections at different focus.
		Field of view is \SI{6}{\micro\metre} (horizontal) by \SI{3}{\micro\metre} (vertical, corresponding to focus).
	{\bf Bottom left:}
		Cross sections from top images, and
		{\bf right:}
		signal-to-noise ratio relative to $m=1$ as a function of interrogation.
		Line denotes a theoretical linear buildup in SNR for reference.
	}
	\label{fig:au_fits}
\end{figure}

\autoref{fig:au_fits} shows multipass images and SNR buildup of a \SI{40}{\nm} gold nanoparticle, with $m=4$ showing a $\SI{\sim10}{\dB}$ increase in signal-to-noise over a single-pass.
The images were fit to a 2D Gaussian, where the fitted volume is taken to be the signal.
The standard deviation of the top-right $15\times\SI{15}{px}$ corner of each image is taken to be the noise.
A similar signal-to-noise improvement of \SI{10}{\dB} would require a ten-fold increase in illumination intensity, or equivalently a ten-fold increase in exposure at a fixed illumination intensity, for a shot-noise limited measurement.
We note that, since we operate near camera saturation, it would not have been possible to achieve a greater single-frame signal by simply increasing the illumination intensity (or camera exposure) at $m=1$;
instead, multiple exposures would need to be acquired, which would reduce throughput and introduce additional readout noise.
Additionally, as we already use a high full well capacity camera, increasing the camera's well depth is not a practical route to achieving higher single-frame signal-to-noise.
Finally, while the performance of our apparatus depends on the specific equipment used (illumination source, camera, etc.), the relative improvement going from $m=1$ to $m=4$ is largely independent of the hardware used.

All preceding images were processed by normalizing to the illumination intensity, obtained by imaging a blank part of the sample:
\begin{eqnarray}
S_\text{image} = \frac{I_\text{raw}-I_\text{dark}}{I_\text{blank}-I_\text{dark}},
\end{eqnarray}
where $S_\text{image}$ is the processed image, $I_\text{raw}$ is the raw image, $I_\text{blank}$ is an image of a blank part of the sample, and $I_\text{dark}$ is an image of dark counts (no illumination).
$I_\text{dark}$ is typically about 500 counts (full well is 2047 counts).

\section{Conclusion}
We have demonstrated a multipass method which is capable of operating at $m=2$ and $m=4$ passes.
Unlike previous work, it is compatible with arbitrary (\emph{e.g.}, CW) illumination sources, and does not require exotic gated cameras.

As the multipass technique works by increasing the scattered signal, the increase in signal-to-noise is agnostic to the source of noise.
It is thus compatible with shot-noise limited detection arising from constrained illumination intensity, acquisition times, or the finite well depth of the detector.
It can also offer an advantage when confronted with technical noise sources such as detector read noise.
It can be used in conjunction with high well depth cameras to further leverage this advantage \cite{welldepth_ieee1,welldepth_ieee2,welldepth_patent,welldepth_adimec}.

The method is compatible with other contrast enhancement methods, \emph{e.g.}, iSCAT or iSCAT-like techniques, by attenuating the reference beam outside of the cavity.
A simple test was carried out with this apparatus by illuminating with a broadband laser and partially blocking the undiffracted (reference) beam in the Fourier plane of the extracavity imaging optics. We observed the expected contrast enhancement.
The multipass method is also compatible with phase contrast (\emph{e.g.}, Zernike) and darkfield modalities, and could also be used in conjunction with techniques utilizing intracavity feedback elements \cite{lowphi}.
For dose sensitive targets, multipass contrast gains can be exploited to reduce sample damage while maintaining image signal-to-noise \cite{theoryslog}.

\section{Acknowledgments}
This work was done as part of the Quantum Electron Microscope collaboration funded by the Gordon and Betty Moore foundation.
B.B.K. would like to thank Adam J. Bowman, Stewart Koppell, and Yonatan Israel for helpful discussions.
\message{The column width is: \the\columnwidth}
\FloatBarrier
\bibliography{main}
\end{document}